\documentclass[journal]{IEEEtran}
\ifCLASSINFOpdf
\else
\fi

\usepackage{amsmath}
\usepackage{lipsum}
\usepackage{graphicx}
\usepackage[justification=centering]{caption}
\usepackage{epstopdf}
\usepackage{tabularx}
\usepackage{algorithm}
\usepackage{algorithmic}
\usepackage{amssymb}

\usepackage{xcolor}

\usepackage[hyphens]{url} 

\begin{document}
\title{Trust2Vec: Large-Scale IoT Trust Management System based on Signed Network Embeddings}

% was not built to handle multiple paragraph
%

\author{Sahraoui Dhelim, Nyothiri Aung, Tahar Kechadi,  Huansheng Ning, Liming Chen and Abderrahmane Lakas

	\thanks{Sahraoui Dhelim, Nyothiri Aung and Tahar Kechadi are with the School of Computer Science, University College Dublin, Ireland.} 
	\thanks{Huansheng Ning is with School of Information Technology and Engineering, Jinzhong University, Shanxi, 030619, China}
    \thanks{Huansheng Ning is also with the School of Computer and Communication Engineering, University of Science and Technology Beijing, Beijing, 100083, China.}
	\thanks{Liming Chen is with the School of Computing, Ulster University, U.K.}
	\thanks{Abderrahmane Lakas is with the College of Information Technology, United Arab Emirates University, UAE}
	\thanks{Sahraoui Dhelim and Nyothiri Aung are co-first authors, they have contributed equally to this work}
	\thanks{Corresponding author: Huansheng Ning (ninghuansheng@ustb.edu.cn).}
}


\markboth{Journal of \LaTeX\ Class Files,~Vol.~14, No.~8, August~2022}%
{Shell \MakeLowercase{\textit{et al.}}: Bare Demo of IEEEtran.cls for IEEE Journals}



\maketitle

\begin{abstract}
  A trust  management system (TMS)  is an integral  component of any  IoT network. A  reliable TMS must guarantee  the network security, data integrity, and act  as a referee that
  promotes legitimate devices, and punishes any  malicious activities. Trust scores assigned by TMSs
  reflect devices' reputations, which can help predict the future behaviours of network entities and
  subsequently judge the reliability of different entities in IoT networks.  Many TMSs have
  been proposed in the literature, these systems are designed for small-scale trust attacks, and can
  deal  with attacks  where  a malicious  device  tries to  undermine TMS  by  spreading fake  trust
  reports. However, these systems  are prone to large-scale trust attacks.  To address this problem,
  in this  paper, we propose  a TMS for large-scale IoT systems  called Trust2Vec, which  can manage
  trust relationships in large-scale IoT systems and can mitigate large-scale trust attacks that are
  performed by hundreds of malicious devices.  Trust2Vec leverages a random-walk network exploration
  algorithm  that  navigates  the  trust  relationship among  devices  and  computes  trust  network
  embeddings, which enables it to analyze the  latent network structure of trust relationships, even
  if there is no  direct trust rating between two malicious devices.  To detect large-scale attacks,
  such as self-promoting  and bad-mouthing,  we propose  a network  embeddings community  detection
  algorithm that detects and blocks communities  of malicious nodes.  The effectiveness of Trust2Vec
  is  validated through  large-scale IoT  network simulation.  The results  show that  Trust2Vec can
  achieve up to 94\% mitigation rate in various network settings.

\end{abstract}


% \fi
\IEEEpeerreviewmaketitle

%@@@@@@@@@@@@@@@@@@@@@@@@@@@@@@@@@@@@@@@@@@@@@@@@@@@@@@@@@@@@@@@@@@@@@@@@@@@@@@
\section{Introduction}

\IEEEPARstart{T}{he} wide  deployment of Internet of  Things (IoT) applications has  created a large
network of interconnected  physical devices, as well  as virtual entities, such  as agents. Managing
trust relationships among  this huge number of IoT  devices is an important part of  IoT security. A
trust  management  system   is  used  to  ensure   network  security  and  data   integrity  in  IoT
\cite{dhelim2021iot}.  A  Trust  management system  (TMS)  can  serve  as  a referee  that  promotes
well-behaved entities and punishes  malicious devices within the network. To do so,  a TMS assigns a
trust score for each entity in the network. A  Trust score is a good indicator for predicting future
behaviors of the network entities and subsequently  judging the reliability of different entities in
an IoT  network. However,  if malicious entities  manage to  alter the trust  scores, the  trust and
reputation indicators might not  reflect the genuine nature of network  entities. Therefore, the TMS
may mistakenly  punish reliable  entities and  reward malicious entities.  Furthermore, such  a fake
trust score  could pose  a serious  threat to  the functioning of  the whole  system and  may enable
network attackers  to gain access  to sensitive information \cite{kumar2021leveraging}.  Trust among
IoT devices is usually measured and  evaluated using two factors \cite{yan2014survey}, namely direct
trust and  indirect trust.  The former  represents the personal  experience of  a given  device with
regards to  other network entities, it  is usually computed  by rating the previous  experience with
these  entities.  The latter  represents  the  reputation  score of  a  device,  it is  computed  by
aggregating multiple ratings given by entities that interacted with the device. 

\begin{figure}[!htbp]
	\centering
	\includegraphics[width=\columnwidth]{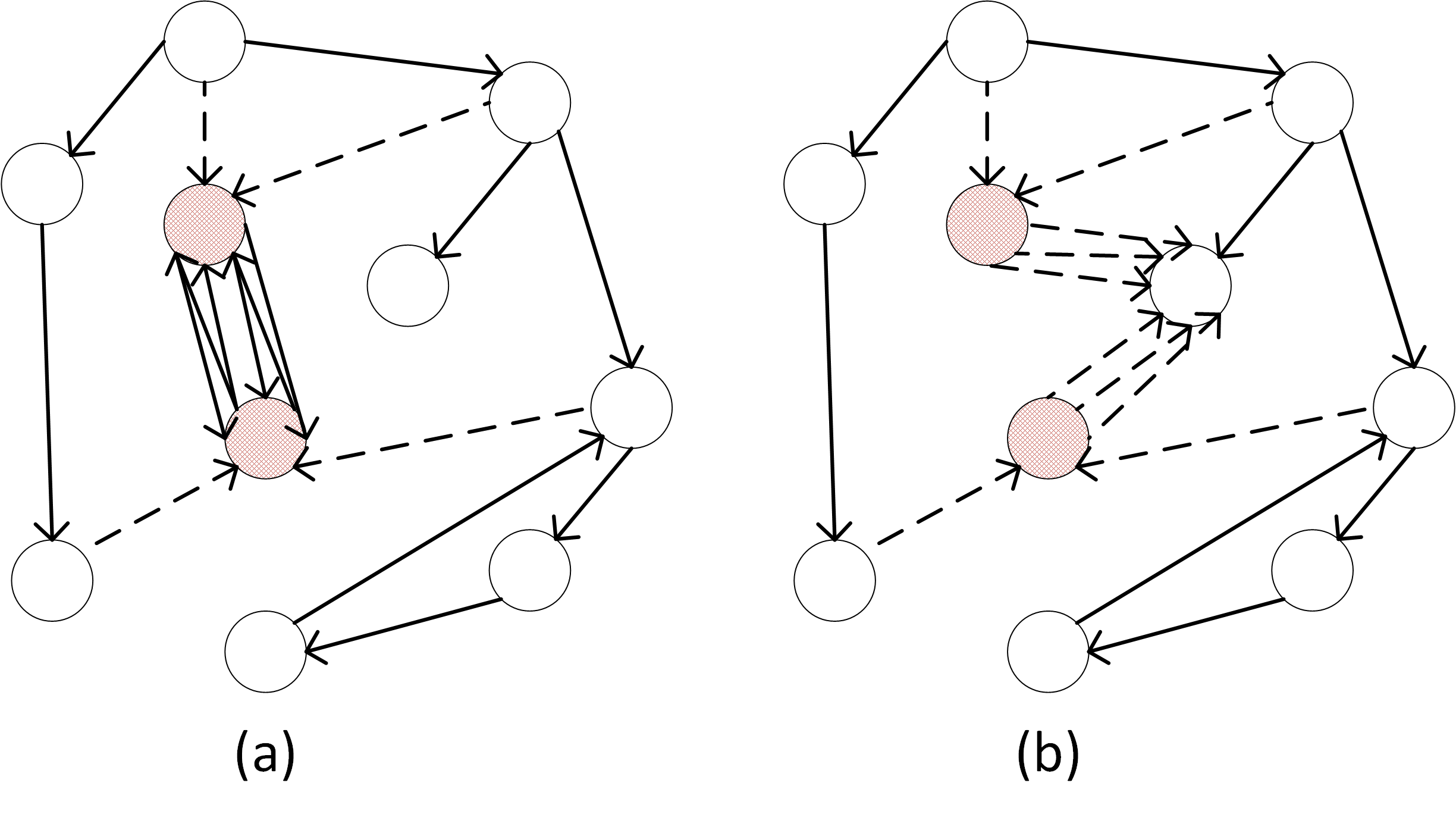}
	\caption{(a) small-scale self-promoting attack. (b) small-scale bad-mouthing attack}
	\label{small_scale}
\end{figure}

\begin{figure}[!htbp]
	\centering
	\includegraphics[width=\columnwidth]{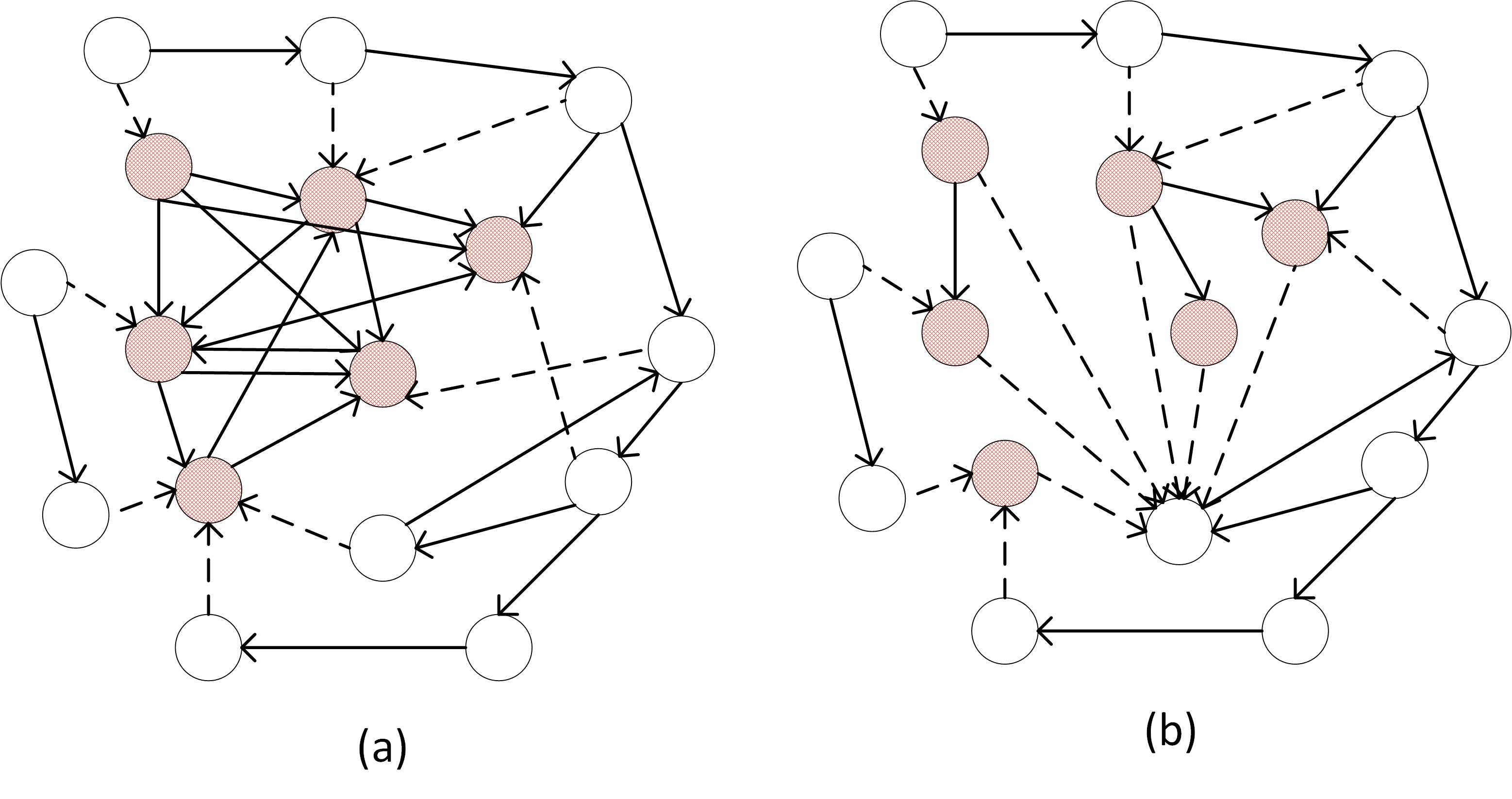}
	\caption{(a) large-scale self-promoting attack. (b) large-scale bad-mouthing attack}
	\label{large_scale}
\end{figure}

The most known  trust-based attacks are self-promoting attack  \cite{kandhoul2019t_cafe} (also known
as  a  good-mouthing  attack)  and bad-mouthing  attack  \cite{kalkan2020trusd}.  In  self-promoting
attacks, malicious devices attempt to illegally increase their trust scores (reputation). The attack
could be conducted by two  nodes, or by a large number of nodes that  work together to achieve their
malicious purpose.  In the most  basic form of  a self-promoting attack,  two nodes provide  a false
report for each  other to promote themselves  as trustworthy entities, hence  increasing their trust
scores (reputation). To mitigate the self-promoting attack,  a TMS must keep track of all previously
reported trust  ratings, and  detect and  punish the entities  that are  involved in  such malicious
activities. In bad-mouthing attack,  attackers usually give bad ratings to a  victim entity in order
to lower its trust score and destroy  its reputation among other nodes.
Figure \ref{small_scale} and
Figure \ref{large_scale} show examples of self-promoting and bad-mouthing attacks. In these figures,
the white circles denote normal entities, and the red circles denote malicious entities that perform
an attack. A solid arrow represents a positive trust rating and a dashed arrow represents a negative
trust rating.  Figure \ref{small_scale}  (a) illustrates an  example of  small-scale self-promoting,
where two  malicious nodes  increase their  trust scores  by repeatedly  giving each  other positive
ratings. Figure \ref{small_scale} (b) demonstrates that two malicious nodes undermine the reputation
of a legitimate node by continuously giving it negative trust ratings. Such small-scale attacks can
be easily mitigated by controlling the rating behaviors  of each entity in the system.  For example,
to prevent  self-promoting attacks, a TMS can limit the number  of positive trust ratings  that two
entities are  allowed to give  to each  other. Similarly, bad-mouthing  attacks can be  prevented by
limiting the number of negative trust ratings that  an entity can assign to another. However, things
get  more complicated  when a  group  of entities  is  collectively involved  in self-promoting  and
bad-mouthing  attacks. For  example, in  Figure  \ref{large_scale} (a)  a group  of malicious  nodes
increase their trust score  by giving each other positive ratings  without attracting any attention,
achieve this in the way that each node gives no more than one positive rating to another node in the
malicious group. Similarly, in Figure \ref{large_scale} (b) a group of malicious nodes performs
bad-mouthing attacks against a normal node by targeting it with unfair ratings.

While  existing  trust  frameworks  can  mitigate  small-scale  trust-related  attacks,  managements frameworks that considered  large-scale trust-related attacks have not been  seen yet. On the other hand, network embedding algorithms have been proven effective when dealing with large-scale graphs and networks \cite{cui2018survey}. Therefore, in this  paper, we  propose a  trust management  framework, dubbed  as Trust2Vec,  for large-scale  IoT systems, which can manage  the trust of millions of IoT devices.  Trust2Vec can mitigate large-scale trust attacks that are performed by hundreds of malicious nodes. Our contributions can be summarized as follows: 

\begin{itemize}
   \item  Propose  a trust  management  framework  that  can  mitigate large-scale  and  small-scale trust-related attacks, such as self-promoting and bad-mouthing attacks. 
   \item  Develop a  random-walk  network algorithm  that navigates  the  trust relationships  among  devices and computes  trust network embeddings. The algorithm enables  the proposed system to analyze the latent network structure of trust relationships. 
    \item We developed a parallelization method for trust attack detection in large-scale IoT systems. Parallelizing made the TMS highly scalable and can manage a large number of network devices with less computational cost.
    
\end{itemize}

The rest of the  paper is organized as follows: Section \ref{sec.2}  reviews existing research about
trust management  in IoT.  Section \ref{sec.3}  describes the  system design  of the  proposed trust
management framework, and how Trust2Vec is used to detect trust-related attacks. Section \ref{sec.4}
presents the  evaluation details and  experiment results. We conclude  the paper and  outline future
research directions in Section \ref{sec.5}.

%@@@@@@@@@@@@@@@@@@@@@@@@@@@@@@@@@@@@@@@@@@@@@@@@@@%%%%%%%%%%%%%%%%%%%@@@@@@@@@@@@@@@@@@@@@@@@@@@@ 
\section{Related Work}
\label{sec.2}
Trust  management  in  IoT  is  a  well-established   research  topic  in  the  literature.  Guo  et
al. \cite{guo2021itcn} proposed a  data collection method for IoT using UAV that  uses a trust score
to  evaluate the  reliability of  data collection  devices. They  concluded that  using trust  as an
evaluation metric for  UAV data collection can  significantly increase the data  accuracy and reduce
data collection costs. Similarly, Liang et al. \cite{liang2021intelligent} investigated the usage of
trust management in UAV-assisted IoT. They proposed  a trust evaluation scheme to identify the trust
of  the mobile  vehicles by  dispatching the  UAV to  obtain the  trust messages  directly from  the
selected devices as evidence. Kumar et  al. \cite{kumar2021tp2sf} introduced a smart city networking
architecture  that  leverages  a  trust   computational  module  to  distinguish  unreliability  and
trustworthiness among  smart city sensors  and devices. Fang  et al \cite{fang2019ddtms}  proposed a
trust management framework, in  which the devices in the cluster start to  detect the nearby devices
within sensing  range, compute  their trust  value, and report  to a  pre-elected cluster  head. The
latter calculates  the aggregated trust  score of each  device in the  cluster. The cluster  head is
periodically re-elected by  the network devices with the cluster.   Chen et al. \cite{chen2018trust}
introduced IoT-HiTrust,  a 3-tier cloud-cloudlet-device hierarchical  trust-based service management
protocol for  large-scale mobile-cloud IoT systems.  Their proposed trust model  combines friendship
similarity, and social  contact similarity to compute  the trust score of network  devices. In their
study, the trust score is represented as a random variable in the range of [0, 1] following the Beta
($\alpha$, $\beta$) distribution. The numbers of positive  and negative experiences of an IoT device
are represented as binomial random variables. They computed  the indirect trust as a weighted sum of
service ratings reported by  other IoT devices, such that trust reports  of socially similar devices
are prioritized. Bahutair et al. \cite{ba2021multi}  introduced a generic trust management framework
that can operate for crowdsourced IoT  services. Their framework leverages a multi-perspective trust
model that obtains the  implicit features of crowd-sourced IoT services.  Each entity is represented
by a set of  characteristics that contribute to the entity's influence on  trust. The trust features
are fed into a machine-learning algorithm that  manages the trust model for crowdsourced services in
an  IoT network.  Marche el  al. \cite{marche2020trust}  discussed possible  trust attacks  that can
affect IoT networks, and introduced a trust  management model that is able to overcome trust-related
attacks.  Specifically, they  proposed  a  decentralized trust  management  model  based on  Machine
Learning algorithms. The model utilized several parameters to compute three trust scores, namely the
goodness, usefulness,  and perseverance  score. Their  model uses these  scores to  detect malicious
nodes  performing trust-related  attacks. Movahedi  et  al. \cite{movahedi2019t}  proposed T-D2D,  a
lightweight trust  model that  evaluates a network  device's trust level  using both  short-term and
long-term evaluation intervals  to mitigate different types of trust-related  attacks. T-D2D records
marginal misbehaving over several successive time slots to reveal the nature of suspicious malicious
nodes with a light misbehaving attitude. To  mitigate bad-mouthing attackers, T-D2D does not rely on
other nodes' recommendations in the case when the  direct trust is not decisive. T-D2D evaluates the
honesty of a recommender  based on the correctness of its recommendations  over time. Ben Abderrahim
et al. \cite{abderrahim2016dtms} introduced DTMS-IoT,  a Dirichlet-based trust management system for
the IoT, which alleviated dishonest trust  recommendations and related attacks by clustering devices
using the k-means algorithm. DTMS-IoT detects IoT  devices' malicious activities, which allows it to
alleviate   the    effect   of    on-off   attacks   and    dishonest   recommendations.  Liu et al. \cite{liu2022semi} introduced a semi-centralized TMS that leverage blockchain for single and multiple domains.  The devices are connected in a centralized fashion and coordinated by a cloud server that manage the rating data ledger, to support cross-domain data exchange the server uses rotation consensus protocol. The proposed TMS aggregates both direct and indirect trust information to compute the trust values of IoT devices. Din et al. \cite{din2021lighttrust}  introduced a  trust framework  for lightweight  devices, which  uses a
centralized  trust authority.  The  framework  manages trust  certificates  that  enable devices  to
exchange  services without  prior  knowledge or  performing trust  computations.  Trust between  two
devices  is  computed  by  direct  observations  in terms  of  delivery  ratio,  compatibility,  and
cooperativeness, while trust recommendations are utilized to determine trust in the case of indirect
observations.  Okuda et  al. \cite{okuda2019community}  proposed a  random-walk community  detection
algorithm that  clusters similar  nodes. Nodes  that frequently appear  when traversing  the network
using finite-length random walk are judged to belong to the same community.  Aung et al \cite{aung2020t,aung2021dynamic} studied trust relationship among driverless cars in the context of vehicular ad-hoc networks (VANET) for route recommendations and path planning, and also trust-based content caching \cite{aung2021vesonet,aung2018accident}. Wu et al. \cite{wu2018information,liao2019security} proposed a deep-learning (DL)-based physical layer authentication scheme which exploits channel state information to enhance the security of mobile edge computing systems. Dhelim et al \cite{dhelim2021survey} studied the trust among social network users for personality-aware recommendation system, they concluded that recommendation accuracy can be significantly improved by adding social factors such as trust and personality traits. Wang et al \cite{wang2021survey} studied trust relationships in human-machine hybrid artificial intelligence. Similarly, Cai et al \cite{cai2020robot} suggested that trust can be established in human-robot interactions.

All  the above-mentioned  trust  frameworks were  designed to  mitigate  small-scale trust  attacks,
without consideration  for large-scale trust attacks.  That is due  to the challenge of  analysing a
large  number  of IoT  devices  with  limited computational  power  required  to analyse  the  trust
relationships. In our proposed  system, we have considered both small-scale,  as well as large-scale
trust attacks.  We have overcome  the computational cost limit  problem by analysing  latent network
embeddings of trust relationships among IoT devices. 

%@@@@@@@@@@@@@@@@@@@@@@@@@@@@@@@@@@@@@@@@@@@@@@@@@@@@@@@@@@@@@@@@@@@@@@@@@@@@@@@@@@@@@@@@@@@@@@@@
\section{System model}
\label{sec.3}
\begin{figure}[!htbp]
	\centering
	\includegraphics[width=\columnwidth]{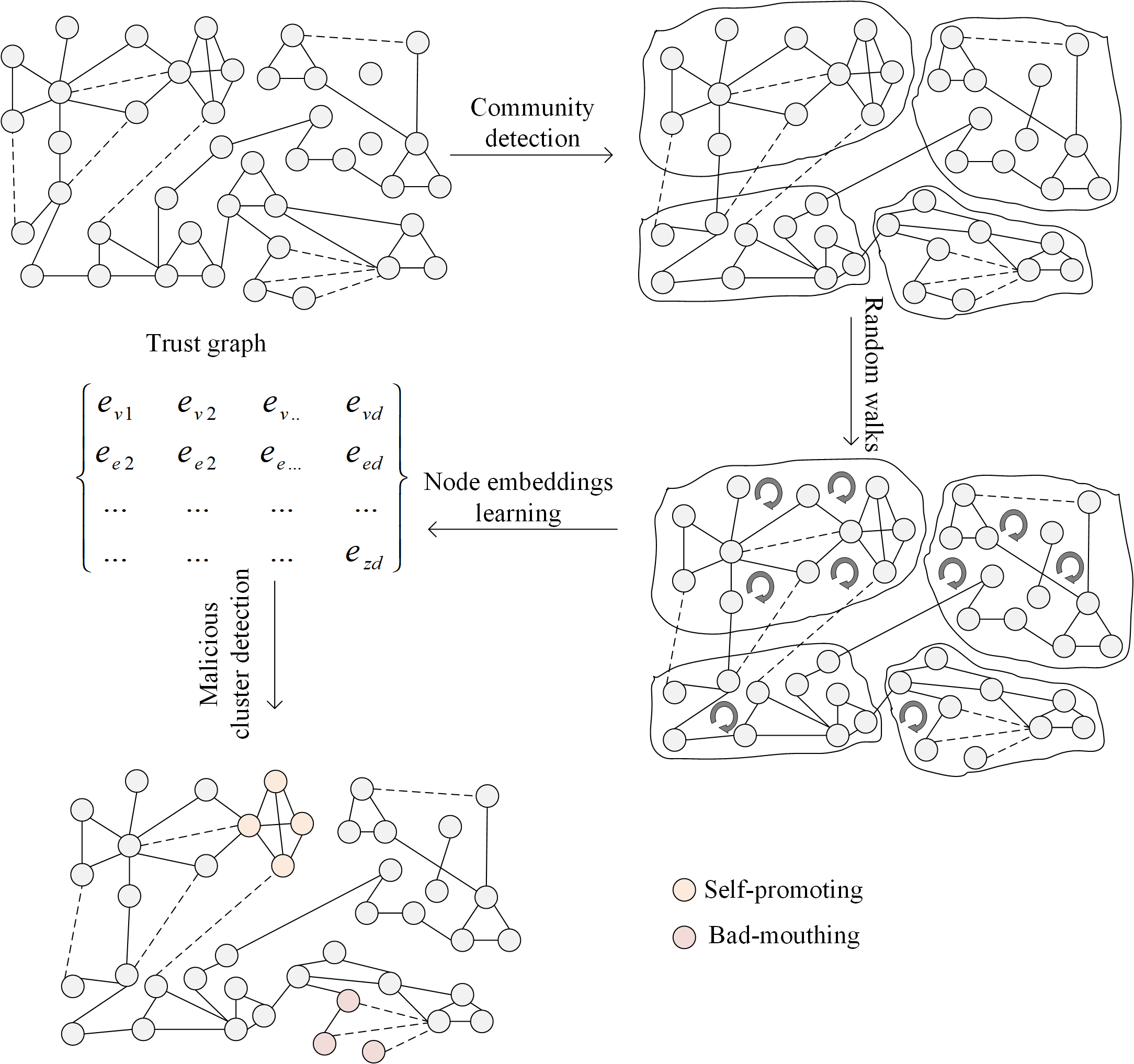}
	\caption{Trust2Vec mitigation scheme phases}
	\label{social_path}
\end{figure}

To detect and mitigate a trust-related attack, Trust2Vec will analyse the network structure of trust
relationships among devices. The main phases of  the attack detection process, as depicted in Figure
\ref{social_path} are: 1)  determine device communities; 2) generate random  walks within each local
community, which yield the devices' trust network embeddings; 3) leverage trust relationship network
embeddings to detect malicious device clusters.

\subsection{Community Detection}

The network of  devices can be structured  as a graph $G=\left(D,T\right)$ that  represents a signed
trust   network  between   devices,   where   the  graph   vertices   represent   the  IoT  devices;
$D=\left\{d_1,d_2,\dots ,d_n\right\}$,  and the edges  represent the previous trust  reports between
graph  nodes; $T=\left\{T^+,T^-\right\}$. An  edge $t_{i,j}\in  T^+$ denotes  that  device $d_j$  is
declared as  trustworthy by device $d_i$,  and $t_{i,j}\in T^-$  denotes that device $d_j$  has been
declared as untrustworthy by  device $d_i$. Given the enormous size of IoT  networks, which can have
millions of  devices, it  is not  viable to  operate on the  overall IoT  network. However,  the IoT
network is generally easy to partition, as it  is composed of smaller IoT subnetworks that are known
as IoT units  \cite{dhelim2021iot}. The edge density  (trust relationships in our  case) within these
IoT units is much higher  than the edge between these Units. This is  based on the observations that
interactions between edge devices to perform edge computing tasks generate a high  number of trust
relationships as  these devices  work together  within the same  local network \cite{naouri2021novel}. This allows us to detect the communities' boundaries without decreasing the resolution limit, an advantage that is not
possible to get with other types of networks, such as social network graphs or user-products graphs \cite{dhelim2020personality}, where the communities tend to overlap with each other \cite{dhelim2020compath}.

\subsection{Random Walks Generation}

To compute the latent network structure, Trust2Vec  generates random walks by navigating the network
through random steps. The  logic behind this approach is that the more  we repeat these random walks
from the same starting device with a fixed length, the more likely the walks will include nodes that
are similar  in terms of network  proximity, either first-order proximity  or second-order proximity
nodes. Let $W_{d_i}(l)$ denotes the random walk starting from device $d_i$ with a walk length $l$. A
random  walk $W_{d_i}(l)$  is a  stochastic process  composed of  a chain  of random  variables
$W_{d_i}\left(l\right)=\left\{w^1_{d_i},w^2_{d_i},\dots  ,w^l_{d_i}\right\}$,  where $w^n_{d_i}$  is
the neighbor  device that  is randomly  chosen among $w^{n-1}_{d_i}$  neighbors.  Relying  on random
walks  to  compute  the  latent  network  structure is  desirable  for  two  reasons.  Firstly,  the
computations  of the  random walks  can be  distributed  among edge  devices, hence  the process  is
distributed and can be performed offline without relying on any remote cloud or server. Secondly, as
the random walks are limited with walk length $l$, the newly added trust relationships among devices
can be easily  accommodated by regenerating random  walks only for the  updated trust relationships,
without  the need  to  recompute  the random  walks  for  the whole  graph. To navigate the  trust
relationships  among devices,  Trust2Vec estimates  the likelihood  of observing  a series  of short
random walks  by adapting  the same  approach in  natural language  modeling, where  the goal  is to
estimate the likelihood of a sentence being present in a corpus. The analogy here is to calculate the trust walk TW that starts from source device $d_o$, and estimate the probability of arriving at destination device $d_d$ given the previously navigated devices through a random walk of length $l$:

\begin{equation}
	\label{eq1}
	{TW\left(d_{s},d_{d},l\right)=Pr\left(d_d\ |\ \left(w^1_{d_s},w^2_{d_s},\dots,w^{l-1}_{d_s}\right)\right)}
\end{equation}

\subsection{Node Embeddings Learning}
The objective  is to  learn the device's latent trust structure  in the  trust graph, not  only the neighboring nodes that  are a few hops away. Trust2Vec computes a device's  trust network structure and represents it as a vector in low dimensional space.  Formally, let $\phi :d\in  D\ \longmapsto
{\mathbb{R}}^{|D|\times  z}$  be the  mapping  function  that  represents  a device's  latent  trust
structure for each device in D. Here z is the length of the vector in the lower dimension, such that $z\ll  \left|D\right|$, hence  it  is easier  to  manipulate  small vectors  rather  than the large adjacency matrix. The likelihood estimation is then represented as:

\begin{equation}
	\label{eq2}
	{Pr\left(d_d\ |\ \left(\phi (w^1_{d_s}),\phi (w^2_{d_s}),\dots, \phi (w^{l-1}_{d_s})\right)\right)}
\end{equation}	

As the random walk length increases, it becomes computationally expensive to calculate the objective
function.    To    overcome   this    problem,    Trust2Vec    leverages   the    skip-gram    model
\cite{mikolov2013distributed},  which inverses  the  problem  by predicting  the  context given  the
missing word  instead of predicting the  word given the  context. The skip-gram model  maximizes the
likelihood of any word  to be observed in the current context without  prior knowledge about current
words. In the context of trust management, Trust2Vec target the following objective function: 

\begin{equation}
	\label{eq3}
	{{\mathop{\mathrm{Min}}_{\phi}\left[-\mathrm{log}Pr\left(\left\{d_{i-w},\dots, d_{i-1}, d_{i+1}, 
	\dots, d_{i+w}\right\}|{\phi (d}_1)\right)\right]}}
\end{equation}

where $w$ is the random walk window size.

Trust2Vec optimizes  the function  (\ref{eq3}) to  build the trust  network representation  for each
device, hence capturing the  latent similarity between network devices. Devices  that have a similar
trust network  structure will have similar  trust vectors in lower-dimensional  space. Measuring the
similarity of devices'  representation in lower-dimensional space allows us  to reveal the community
membership, thus detecting malicious devices that perform trust-based attacks.

To computing devices' trust relationships and extract trust network embeddings, we developed Algorithm \ref{deviceTrustAlgo}. The algorithm takes as an input the graph of devices in the studied edge environment, and generate the matrix $\phi \in {\mathbb{R}}^{|D|\times z}$ that represents the trust relationship network structure in lower dimensional space. Firstly, the algorithm generates the random walks of length $l$ for $\mathrm{\lambda}$ times starting from each device $d_x$ (line 3-5), and for each device within the $\omega $ hop away in the random walk $W_{\mathrm{\ }d_x}$ apply the SkipGram model to map every device $d_y$ to its representation vector ${\phi (d}_y)\in {\mathbb{R}}^z$. Given the low dimension representation, we are aiming to maximize the probability of the device neighbors in random walks (line 7). The posterior probability can be computed using basic classifiers such as logistic regression. However, this approach is not feasible as the number of devices increase, and become computational expensive to perform. To address this, we leverage hierarchical softmax \cite{perozzi2014deepwalk}, which maps all the network nodes (devices in our case) to a binary tree. In this way, the prediction problem is pivoted to maximizing the probability of path navigation from the root to the leaf of the tree that identify that device. In case a device $d_k$ is defined by the sequence $\left(b_0,b_1,\dots .,b_{\left\lceil log\left[D\right]\right\rceil }\right)$ where $b_0$ is the root and $b_{\left\lceil log\left|D\right|\right\rceil }=d_k$ then eq(\ref{eq3}) can be computed using binary classifier, which reduces the complexity of computing $Pr(d_k|\phi (d_y))\ $ from $O(\left|D\right|)$ to $O(log\left|D\right|)$ as showed bellow in eq(\ref{eq4})

\begin{equation}
	\label{eq4}
	{{\mathrm{Pr} \left(d_k\mathrel{\left|\vphantom{d_k {\phi (d}_y)}\right.\kern-\nulldelimiterspace}{\phi (d}_y)\right)\ }=\prod^{\left\lceil log\left|D\right|\right\rceil }_{l=1}{Pr(d_k|\phi (d_y))}}
\end{equation}

To optimize eq(4), Trust2Vec utilizes Stochastic gradient descent (SGD). The derivatives are computed using the back-propagation algorithm, SGD learning rate is initialized as 2.5\% at the start of the training and then decreased linearly with the count of devices encountered so far.

\begin{algorithm}
	\caption{Device\_Trust\_Embeddings}
	\label{deviceTrustAlgo}
	\begin{flushleft}
	\textbf{Input}\newline 
	graph $G=\left(D,T\right)$\newline
	Walk window size $\omega $\newline  
	Low dimension vector size $z$\newline  
	Random walks count per device $\mathrm{\lambda }$\newline  
	Random walks length $l$\newline
	\textbf{Output} \newline
	$\phi \in {\mathbb{R}}^{|D|\times z}$ Matrix of device trust embeddings
	\end{flushleft}
	\begin{algorithmic}[1]
	\STATE Sample\_matrix($\phi ,D)$
	\STATE Generate\_Binary\_Tree(T,D)
	\FOR {$i:\ 0\to \mathrm{\ }\mathrm{\lambda }$}
	\FORALL{$d_x\in D$}
	\STATE $W_{\mathrm{\ }d_x}=RandomWalk(G,\mathrm{\ }d_x,l)$
	\FORALL{$d_y$ in $W_{\mathrm{\ }d_x}$}
	\STATE $J\left(\phi \right)=\ -{\mathrm{log} Pr\left(d_k|\phi (d_y)\right)\ }$
	\ENDFOR
	\ENDFOR
	\ENDFOR
	\end{algorithmic} 
\end{algorithm}

\subsection{Trust attack detection}

After computing the low dimensional trust network structure using Trust2Vec, the resulting embeddings are used to detect trust attacks. We propose an algorithm which can detect large-scale bad-mouthing and self-promoting attacks, as shown in Algorithm \ref{trust_attack_algo}. Given the network embeddings of trust graph $\phi (G)$, the source device $d_s$ that reports the trust relationship (trustor), and the destination device $d_d$ that received trust level (trustee). In case of a positive trust report, it is checked for possible large-scale self-promoting. Lines 1-8 checks if the similarity of embeddings vectors of the trustor and trustee is greater than the embedding similarity threshold $\alpha $, the set of suspected devices is denoted as ${\mathrm{\Omega }}_P$, which is the union of the previous positive trustees and trustors of the trustor device $d_s$. The set of malicious self-promoting cluster ${\mathrm{M}}_s$ is determined by comparing the positive trust report to devices within ${\mathrm{\Omega }}_P$, and to devices outside ${\mathrm{\Omega }}_P$, which are denoted as $\overline{{\mathrm{\Omega }}_P}$, and classified as malicious nodes if the cardinality difference is greater than the self-promoting similarity threshold $\beta$. In the case of a negative trust report, it is checked against bad-mouthing attack as shown in lines 9-20. The set of suspected device is denoted as ${\mathrm{\Omega }}_N$, which contain the devices $N_{in}(d_d)$ that previously given negative trust report against device $d_d$. If two or more devices within ${\mathrm{\Omega }}_N$ have low dimension similarity greater than the bad-mouthing similarity threshold $\gamma $, then these nodes are classified as malicious bad-mouthing community ${\mathrm{M}}_b$.

\begin{algorithm}
	\caption{Trust\_Attack\_Detection}
	\label{trust_attack_algo}
	\begin{flushleft}
	\textbf{Input}\newline 
	Network embeddings of trust graph $\phi (G)$\newline 
	$d_s$ trustor device\newline 
	$d_d$ trustee device\newline
	\textbf{Output} \newline
	${\mathrm{M}}_s$ malicious self-promoting community\newline
	${\mathrm{M}}_b$ malicious bad-mouthing community 
	\end{flushleft}
	\begin{algorithmic}[1]
	\IF{($R\left(d_s,d_d\right)>0$)}
	\IF{(Sim($\phi \left(d_s\right),\phi (d_d)$)$>\alpha $)}
	\STATE ${\mathrm{\Omega }}_P\mathrm{\leftarrow}P_{in}(d_s)\cup P_{out}(d_s)$
	\FORALL{$d_i\in {\mathrm{\Omega }}_P$}
	\IF{$(\left|P_{out}\left(d_i,{\mathrm{\Omega }}_P\right)\right|-\left|P_{out}(d_i,\overline{{\mathrm{\Omega }}_P})\right|)>\beta $}
	\STATE ${\mathrm{M}}_s\mathrm{\leftarrow }{\mathrm{M}}_s\mathrm{\cup }\left\{d_i\right\}$
	\ENDIF
	\ENDFOR
	\ELSE
	\STATE ${\mathrm{\Omega }}_N\mathrm{\leftarrow }N_{in}(d_d)$
	\FORALL{$d_i\in {\mathrm{\Omega }}_N$}
	\IF{$(\left|N_{out}(d_i,d_d)\right|)$}
	\FORALL{$d_j\in {\mathrm{\Omega }}_N-\left\{d_i\right\}$}
	\IF{$(Sim\left(\phi \left(d_s\right),\phi \left(d_d\right)\right)>\gamma )$}
	\STATE ${\mathrm{M}}_b\mathrm{\leftarrow }{\mathrm{M}}_b\mathrm{\cup }\left\{d_i,d_j\right\}$
	\ENDIF
	\ENDFOR
	\ENDIF
	\ENDFOR
	\ENDIF
	\ENDIF
	\end{algorithmic} 
\end{algorithm}

\section{Evaluation }
\label{sec.4}

\subsection{Evaluation baselines}

To test the effectiveness of the proposed system, we have compared its performance with the following trust management systems from the literature.

DDTMS \cite{fang2019ddtms}: In this system, the devices in the cluster start to detect the nearby devices within sensing range, and compute their trust value, and report that to a pre-elected cluster head. The latter calculates the aggregated trust score of each device in the cluster. The cluster head is periodically reelected by the network devices located within the cluster. In this system, the trust value is computed as: $Tij\mathrm{=}\alpha \mathrm{\times }DTij\mathrm{+}\beta \mathrm{\times }RTj$. Whereas, $\alpha$ is the weight of weight and $\beta$ is the weight of indirect observation, such that $\alpha$+$\beta$=1.

T-D2D \cite{movahedi2019t}: In this system, the overall trust level is computed by aggregating the direct trust level that account for the direct interaction between the two devices, and indirect trust that rely on other devices' recommendations. The total trust level between device i and device j is calculated as: $TTL_{i,j}=(1-\omega) DTL_{i,j}+ \omega ITL_{i,j}$ , where $DTL_{i,j}$ denotes direct trust level between device i and device j, $d$ denotes inderct trust level between device i and device j, and $\omega$ is the attention factor.

Liu-Trust: is a semi-centralized TMS that leverage blockchain for trust score management. The total trust value of device $i$ in device $j$ is denoted by $T^j_i\mathrm{(}t\mathrm{)=}\alpha \mathrm{\times }{DT}^j_i\mathrm{(}t\mathrm{)+}\beta \mathrm{\times }{IT}^j_i\mathrm{(}t\mathrm{)}$ , which is computed by aggregating direct trust value ${DT}^j_i\mathrm{(}t\mathrm{)}$, as well as indirect trust value ${IT}^j_i\mathrm{(}t\mathrm{)}$. The direct trust is computed as ${DT}^j_i\left(t\right)\mathrm{=}{ResT}^j_i\left(t\right)\mathrm{\times }RaT^j_i\mathrm{(}t\mathrm{)}$, where ${ResT}^j_i\left(t\right)$ is the response trust of device $i$ to device $j$, and is defined as the probability of device $i$ in whether device $j$ can provide information on time, and $RaT^j_i\mathrm{(}t\mathrm{)}$ is the rating trust and it represents the trustworthiness of device $i$ about another device $j$ on the aspect that device $j$ can provide reliable data on the request of $i$.

LightTrust: is a light-weight TMS, in which the trust between two nodes p and q is estimated by aggregating: device compatibility ${com}_{p\mathrm{\to }q}$, cooperativeness ${coop}_{p\mathrm{\to }q}$, and delivery ratio ${dl}_{p\mathrm{\to }q}$ as shown in eq (x), in addition to experience and previous knowledge. 
\[T\left(p,q\right)=\sum{({com}_{p\to q},{coop}_{p\to q},\ {dl}_{p\to q})}\] 

DTMS-IoT \cite{abderrahim2016dtms}: is a TMS that alleviate dishonest trust recommendations and related attacks by clustering the devices using k-means algorithm.

\subsection{Experiments}
\begin{figure*}[htb]
	\centering
	\includegraphics[ scale=0.35]{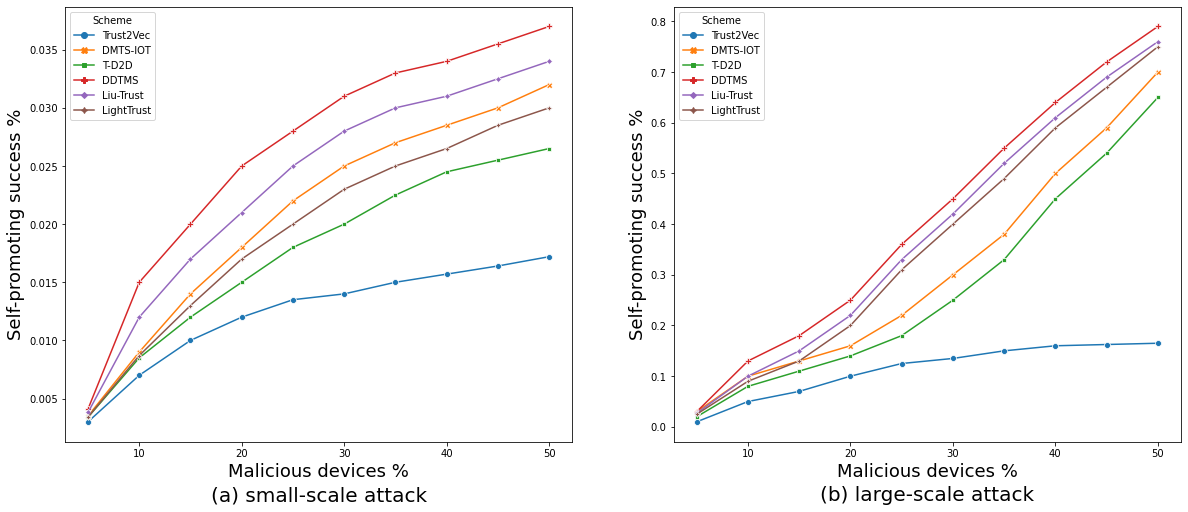}
	\caption{Self-promoting in various malicious devices count}
	\label{result1}
\end{figure*}

\begin{table}
	\centering
	\caption{Simulation parameters}
	\label{simulation_parameters}
	\begin{tabular}{|l|l|l}
		\cline{1-2}
		\textbf{Parameter}   & \textbf{Value}   \\ \cline{1-2}
		OMNet++ & V5.7.0   \\ \cline{1-2}
		INET & V3.7.1             \\ \cline{1-2}
		Mobility Type & Linear Mobility \\ \cline{1-2}
        Mobility speed  & 10mps \\ \cline{1-2}
        Update Interval & 100ms \\ \cline{1-2}
        Transmitter power & 3.5mW \\ \cline{1-2}
	\end{tabular}
\end{table}

We have compared the proposed system with the above-mentioned baselines in various scenario and experiment settings. The simulation is performed using INET, an open-source model library framework of OMNet++ simulator \cite{varga2010omnet}, that can simulate wired and wireless networks, and also support mobility module that can be used to simulated IoT and Fog/Edge computing networks. Table \ref{simulation_parameters} shows the simulation environment details. We simulate the network with different number of devices and malicious devices percentages to observe the effect on the overall performance. We simulate 10000 devices, The mobile devices are randomly placed at the beginning of simulation, and moves according to INET's linear mobility model. The trust scores of the devices within each cluster are stored in a local fog server that is usually the gateway to the external network. The trust value of each device is initialized as 0, and can vary between -1 and 1. Devices can express their trust level regarding their neighbouring device following device to device interaction such as data exchange interactions, or common computational or data offloading tasks. For small-scale attacks, we randomly choose a device that tries to self-promote or bad-mouth one of its neighbors that is randomly selected, we repeat the attack until the number of fake trust report submitted by the attackers represent a certain percentage (attack density) of the total submitted trust reports. We evaluate the system's performance with difference attack densities (5\% to 50\%). We also evaluate the system with different malicious devices percentages, in which we randomly select a certain percentage of devices to perform self-promoting or bad-mouthing attacks, we have simulated the system in different malicious devices percentage settings (5\% to 50\%). For large-scale attacks, we randomly select x (depending on the malicious devices percentage) devices as a group malicious devices that self-promote each others, in which each device iterate and self-promote all the other devices within the malicious group, as shown in the example in Figure  \ref{large_scale} (a). for bad-mouthing, we randomly select a victim device, all the devices in the malicious group will bad-mouth the victim device, as shown in the example in Figure  \ref{large_scale} (b).

We evaluate the proposed system and other baselines based on the following metrics: (1) Self-promoting attack resilience: The ability to accurately detect small-scale and large-scale self-promoting attacks without mistakenly blocking legitimate devices. (2) Bad-mouthing attack resilience: The ability to identify small-scale and large-scale bad-mouthing attacks without mistakenly blocking legitimate devices. The attack success rate is defined as the ratio of succeed attacks (e.g. self-promoting trust report) from all attempted attacks by all nodes in the network, as defined in eq (\ref{attack_success_rate}), where $s$ is the number of simulated devices, $AS_i$ is the total succeed attacks by device $i$, and $AA_i$ is the total attack attempts by device $i$.

\begin{equation}
\label{attack_success_rate}
{ASR=\frac{\sum_{i=1}^{s}\frac{AS_i}{AA_i}}{s}}
\end{equation}
\begin{figure*}[hbt]
	\centering
	\includegraphics[scale=0.35]{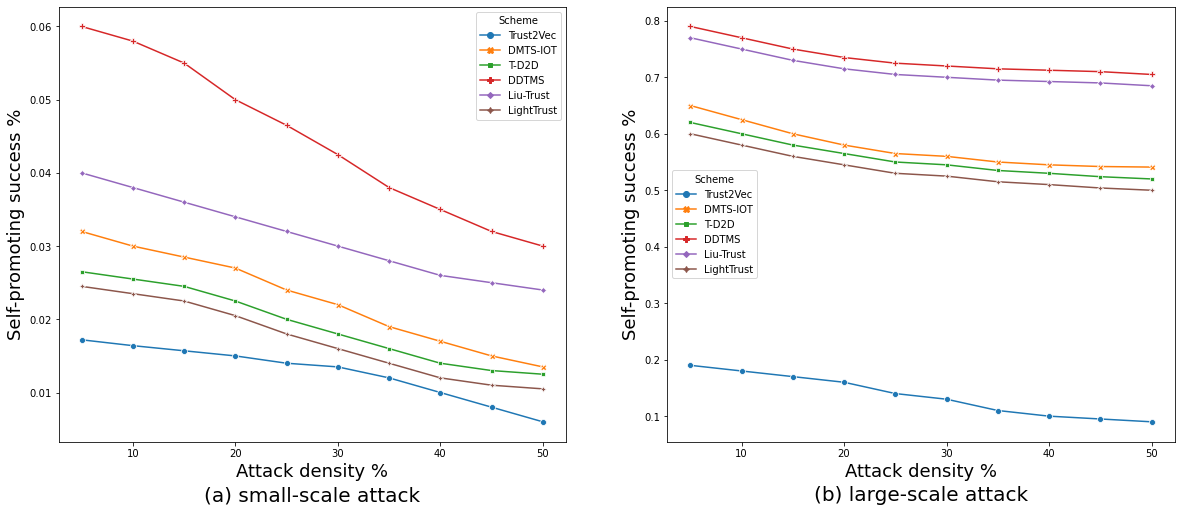}
	\caption{Self-promoting in various attack densities}
	\label{result2}
\end{figure*}

\begin{figure*}[ht]
	\centering
	\includegraphics[scale=0.35]{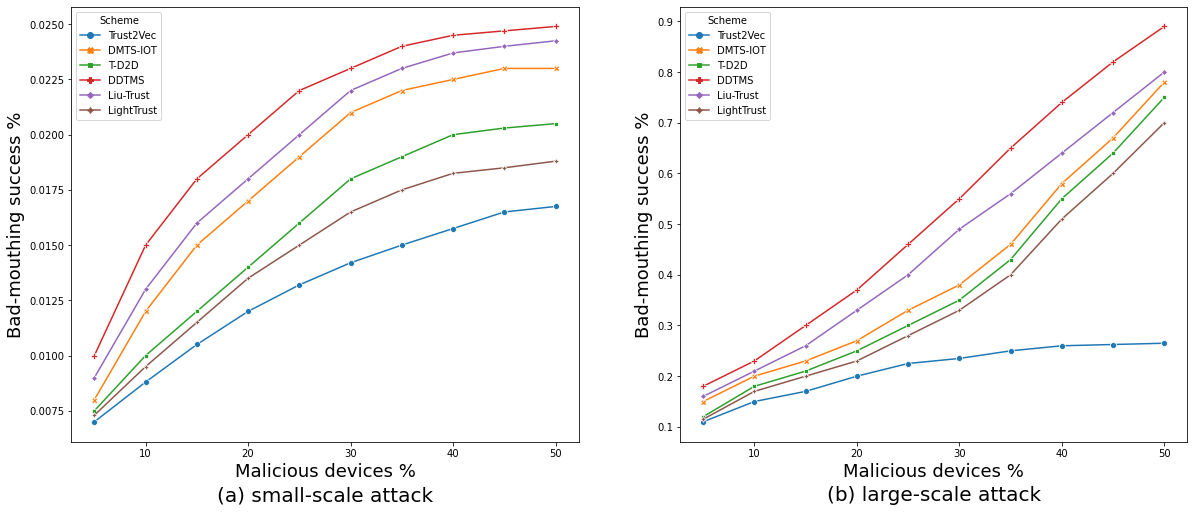}
	\caption{Bad-Mouthing in various malicious devices count}
	\label{result3}
\end{figure*}

\begin{figure*}[ht]
	\centering
	\includegraphics[scale=0.35]{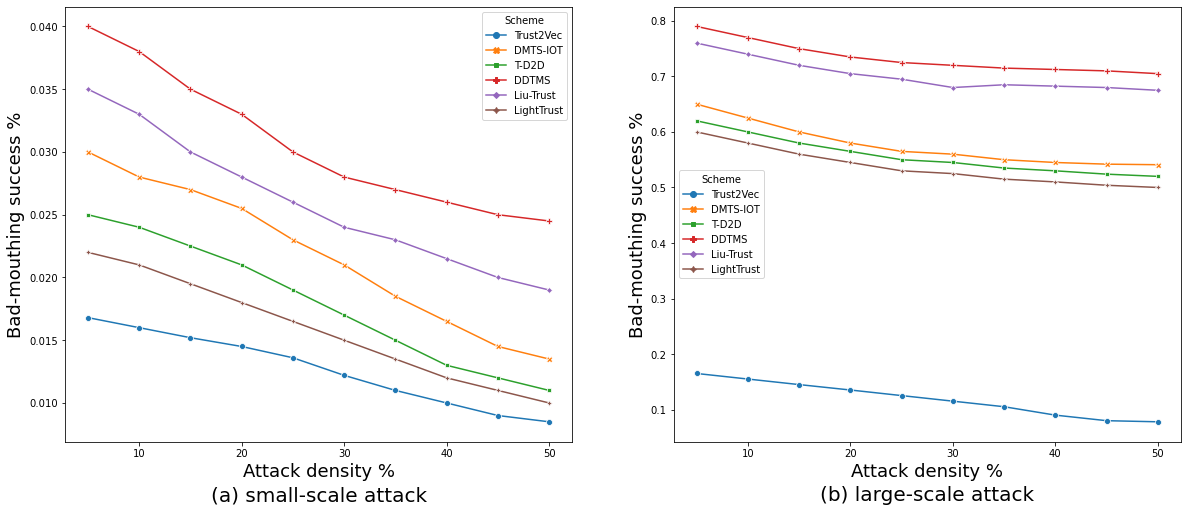}
	\caption{Bad-Mouthing in various attack densities}
	\label{result4}
\end{figure*}

\subsection{Results and discussion}

Figure \ref{result1} shows self-promoting success rate with different malicious devices percentage from 5\% to 50\%. In Figure \ref{result1} (a) the attack is performed as small-scale self-promotion, where the malicious devices are randomly chosen, and each two malicious devices try to inflate each other's trust value by broadcasting fake trust reports to other nodes in network. Figure \ref{result1} (b) displays a large-scale attack, where randomly selected malicious devices inflate their trust scores by distributing trust reports among their group rather than through multiple mutual trust reports, hence avoid being detected. From Figure \ref{result1}(a), we can observe that the percentage of a successful self-promoting attack increases proportional with malicious devices percentage for all studied systems. All baselines have relatively low attack success rate, with 0.005\% when there is  5\% of the devices are malicious. The attack success rate increase when more devices malicious participate in the attack, as it become more difficult to distinguish legitimate trust ratings from self-promoting ratings; however, the attack is mitigated as the attackers are stopped once detected, and the attack success rate stabilizes with less than 0.03 for all baselines, except DDTMS and Liu-Trust, they fail to detect malicious nodes as they rely on neighboring devices' observations, which can be misleading if the neighbors are among the malicious nodes. Unlike Figure \ref{result1}(a) where Trust2Vec has similar performance with the studied baselines, the upper hand of Trust2Vec is obvious in Figure \ref{result1}(b) which shows the success rate of large-scale self-promoting attack with different malicious devices percentage from 5\% to 50\%. We can observe that Trust2Vec is the only system that can mitigate attacks as the percentage of malicious devices increases. That is because Trust2Vec analyses not only the direct trust link but also the latent trust graph structure, whereas other baselines focus on direct trust links between devices. 

Figure \ref{result2} shows self-promoting success rate with different attack density. The higher the attack density, the more malicious devices attempt to inflate their trust score. Figure \ref{result2} (a) and (b) show small-scale  and large-scale attack scenario. From Figure \ref{result2} (a), we can observe that all the attack success rate plummets when the attack density increases, that is because in high attack density, malicious device become more aggressive by sending successive fake trust reports, hence they are easily detected and eventually blocked. In Figure \ref{result2} (b), we can easily observe that Trust2Vec copes well with increase of attack density in large-scale scenarios, unlike other baselines that could not mitigate large-scale attacks.

Figure \ref{result3} shows bad-mouthing success rate with different malicious devices percentage from 5\% to 50\%. Figure \ref{result3} (a) and (b) display small-scale and large-scale attacks respectively. As seen in self-promoting attacks, the studied baselines have similar performance in small-scale settings. Nonetheless, the superiority of Trust2vec is obvious in large-scale settings, that is because the bad-mouthing is traced back by analysing the malicious devices' latent trust network structure through network embedding comparison, unlike the studied baselines that rely solely on the direct observation of neighboring devices.

Figure \ref{result4} shows bad-mouthing success rate with different attack density. With small-scale attack in Figure \ref{result4} (a), we can observe that all the studied baselines perform better when with higher attack density, with 0.04\% attack success rate at worst (DDTMS) when the attack is at 5\% density. As the attack density increases, it become much easier to detect and block malicious nodes. However, in large-scale attack scenario shown in Figure \ref{result4} (b), all baselines (DDMTS, T-D2D, Liu-Trust, LightTrust and DMTS-IOT) fail to detect malicious nodes, with at least 55\% attack success rate in all attack densities. Nevertheless, Trust2Vec was able to mitigate large-scale by 82\% in less dense attacks and up to 90\% percent in highly dense attacks.

\section{Conclusion}
\label{sec.5}

In this paper we have proposed a trust management system for large-scale IoT systems named Trust2Vec. Unlike state-of-the-art trust frameworks that focus only on small-scale IoT networks, Trust2Vec can be leveraged to manage trust relationships among devices in large-scale IoT application. Trust2Vec had been validated through large-scale IoT network simulation. The results show that Trust2Vec can achieve up to 94\% mitigation rate in various network scenarios. 
The proposed trust management system can be further improved from various aspects:
\begin{itemize}
    \item The proposed system focus on general IoT applications, where the devices can include fixed devices such as sensors and mobile devices such as mobile phones. However, things may differ in high dynamic environments such as vehicular network. Extending the proposed system to be customized for scenario-specific IoT applications is one of our future directions
    \item Trust2Vec can be extended to manage trust in of virtual network entities  by a software defined network.
    \item The proposed system manages trust scores of network devices. In our next work, we will extend that to include trust management of data entities as well.
\end{itemize}

\section*{Acknowledgment}
This work was supported by:

The Insight Centre for Data Analytics funded by Science Foundation Ireland under Grant Number 12/RC/2289 P2.

CONSUS project funded by the SFI Strategic Partnerships Programme (16/SPP/3296) and is co-funded by Origin Enterprises Plc.

The National Natural Science Foundation of China under Grant 61872038.

\ifCLASSOPTIONcaptionsoff
  \newpage
\fi



%\bibliographystyle{IEEEtran}

\bibliographystyle{IEEEtran}

\bibliography{refs}

\vskip 0pt plus -1fil

\begin{IEEEbiography}[{\includegraphics[width=1in,height=1.25in,clip,keepaspectratio]{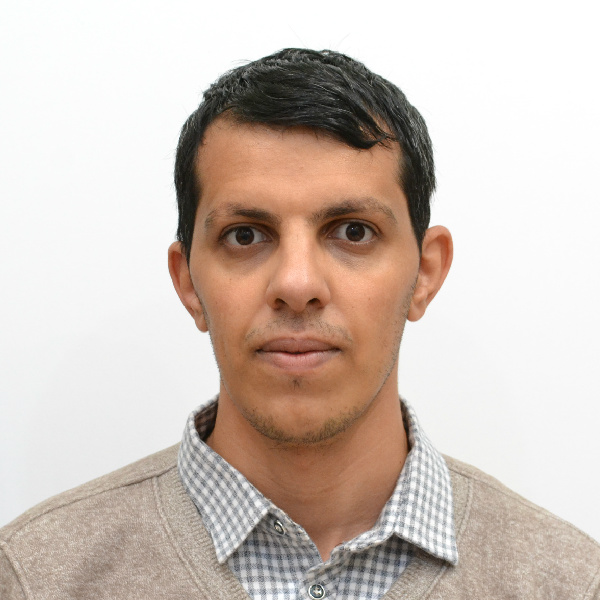}}]{Sahraoui Dhelim}
	is a postdoctoral researcher at University College Dublin, Ireland. He was a visiting researcher at Ulster University, UK (2020-2021). He obtained his PhD degree in Computer Science and Technology from the University of Science and Technology Beijing, China, in 2020. And a Master's degree in Networking and Distributed Systems from the University of Laghouat, Algeria, in 2014. He serves as workshop chair of Cyberspace congress (CyberCon). His research interests include Social Computing, Digital Agriculture, Deep-learning, Recommendation Systems and Intelligent Transportation Systems.
\end{IEEEbiography}

\vskip 0pt plus -1fil
\begin{IEEEbiography}[{\includegraphics[width=1in,height=1.25in,clip,keepaspectratio]{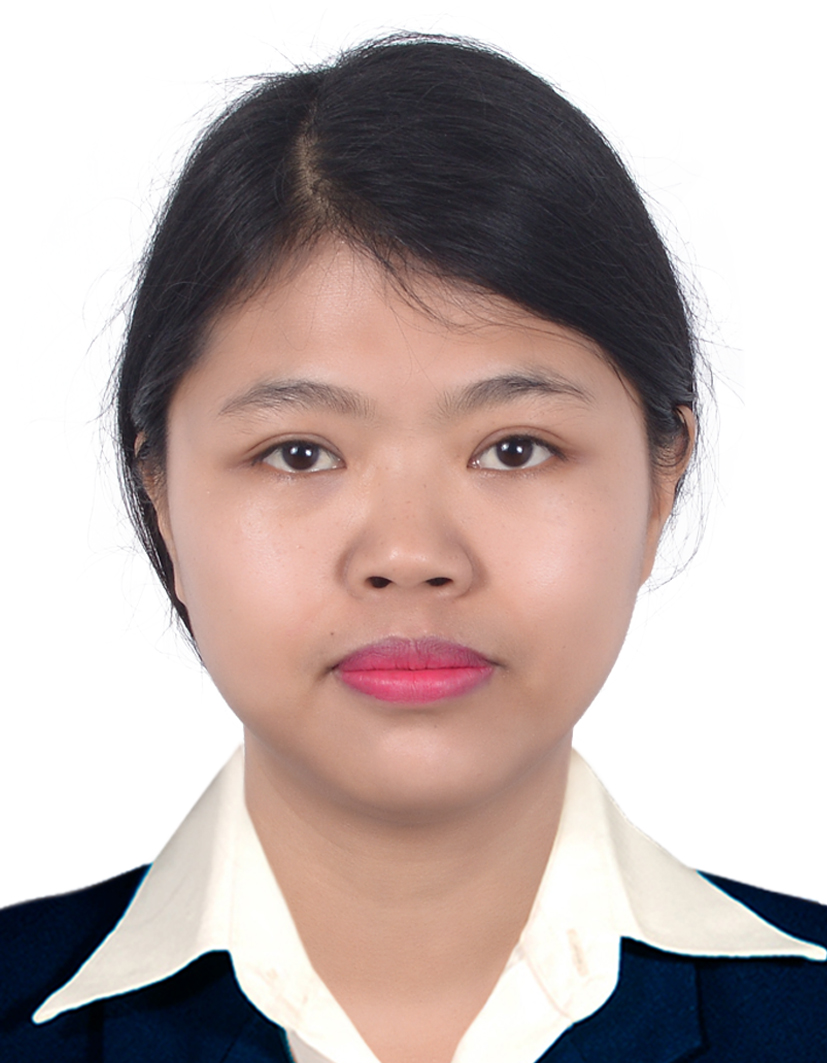}}]{Nyothiri Aung}
is a postdoctoral researcher at University College Dublin, Ireland. She received her PhD in Computer Science and Technology from University of Science and Technology Beijing, China, 2020. And a Master's of Information Technology from Mandalay Technological University, Myanmar, 2012. She worked as a tutor at the Department of Information Technology in Technological University of Meiktila, Myanmar (2008-2010). And System Analyst of ACE Data System, Myanmar (2012-2015). Her research interests include Social Computing, Medical image analysis, and Intelligent Transportation Systems.
\end{IEEEbiography}

\vskip 0pt plus -1fil
\begin{IEEEbiography}[{\includegraphics[width=1in,height=1.25in,clip,keepaspectratio]{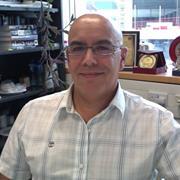}}]{Mohand Tahar Kechadi}
is a full professor in school of computer science, University College Dublin, Ireland. He received master’s and Ph.D. degrees in computer science from the University of Lille 1, France. His research interests include data mining, distributed data mining heterogeneous distributed systems, grid and cloud computing, and digital forensics and cyber-crime investigations. He is a member of the Communications of the ACM journal and IEEE Computer Society. He is an Editorial Board Member of journal of Future Generation Computer Systems.
\end{IEEEbiography}

\vskip 0pt plus -1fil

\begin{IEEEbiography}[{\includegraphics[width=1in,height=1.25in,clip,keepaspectratio]{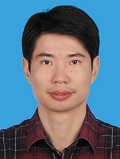}}]{Huansheng Ning}
	Received his B.S. degree from Anhui University in 1996 and his Ph.D. degree from Beihang University in 2001. Now, he is a professor and vice dean of the School of Computer and Communication Engineering, University of Science and Technology Beijing, China. His current research focuses on the Internet of Things and general cyberspace.
	He is the founder and chair of the Cyberspace and Cybermatics International Science and Technology Cooperation Base.
	He has presided many research projects including Natural Science Foundation of China, National High Technology Research and Development Program of China (863 Project). He has published more than 150 journal/conference papers, and authored 5 books. He serves as an associate editor of IEEE Systems Journal (2013-Now), IEEE Internet of Things Journal (2014-2018), and as steering committee member of IEEE Internet of Things Journal (2016-Now).
\end{IEEEbiography}

\vskip 0pt plus -1fil

\begin{IEEEbiography}[{\includegraphics[width=1in,height=1.25in,clip,keepaspectratio]{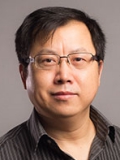}}]{Liming Chen}
	is a professor in the School of Computer  Science  and  Informatics  at  University  of  Ulster,  Newtownabbey,  United  Kingdom.  He  received his  B.Eng  and  M.Eng  from  Beijing  Institute  of Technology  (BIT),  Beijing,  China,  and  his  Ph.D  in Artificial Intelligence from De Montfort University,UK.  His  research  interests  include  data  analysis,ubiquitous computing, and human-computer interaction. Liming is a Fellow of IET, a Senior Member of IEEE, a Member of the IEEE Computational Intelligence Society (IEEE CIS), a Member of the IEEE CIS Smart World Technical Committee (SWTC), and the Founding Chair of the IEEE CIS SWTC Task Force on User-centred Smart Systems (TF-UCSS). He has served as an expert assessor, panel member and evaluator for UK EPSRC (Engineering and Physical Sciences Research Council, member of the Peer Review College), ESRC (Economic and Social Science Research Council), European Commission Horizon 2020 Research Program, Danish Agency for Science and Higher Education, Denmark, Canada Foundation for Innovation (CFI), Canada, Chilean National Science and Technology Commission (CONICYT), Chile, and NWO (The Netherlands Organisation for Scientific Research), Netherlands.
\end{IEEEbiography}

\vskip 0pt plus -1fil

\begin{IEEEbiography}[{\includegraphics[width=1in,height=1.25in,clip,keepaspectratio]{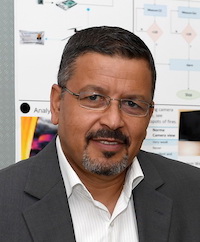}}]{Abderrahmane Lakas}
is a Professor at the Computer and Network Engineering department in the College of IT at UAE University. He holds an MS, and PhD in Computer Systems from the University of Pierre et Marie Curie (Paris VI, France). He has several years of both academic and industrial experience. He spent two years as a postdoc researcher at School of Computing and Communication at the University of Lancaster in UK. He is the head of CAST (Connected Intelligent Autonomous Systems) research group and the Connected Autonomous Intelligent Systems Lab (ASIL). Prior to joining UAE University, he held several industrial positions in several companies in Canada and the US including at Netrake (Plano, TX, USA), Nortel Networks (Ottawa, Canada), and Newbridge (Ottawa, Canada). His current research interests include intelligent transportation systems, vehicular ad hoc networks, unmanned ground and aerial vehicles, autonomous systems, smart cities, Internet of Things, and QoS. Dr. Lakas has published several research papers in scholarly journals. He is member of the TPC and reviewer of several renown conferences and serves in the editorial board of few journals.
\end{IEEEbiography}


%\vfill


\end{document}